\documentclass[12pt]{iopart}  
\usepackage{iopams}  

\usepackage{graphicx}
\usepackage{epsfig}
\usepackage{epstopdf}
\newcommand{\lt}{\left(}
\newcommand{\rt}{\right)}
\newcommand{\lqu}{\left[}
\newcommand{\rqu}{\right]}
\newcommand{\lgr}{\left\{}
\newcommand{\rgr}{\right\}}

\newcommand{\be}{\begin{equation}}
\newcommand{\ee}{\end{equation}}
\newcommand{\ba}{\begin{eqnarray}}
\newcommand{\ea}{\end{eqnarray}}
\newcommand{\fr}{\frac}
\newcommand{\nn}{\nonumber}
\newcommand{\se}{\section}
\newcommand{\sse}{\subsection}

\newcommand{\R}{\mathbb{R}}
\begin{document}

%-------------------------------------------------------------------------------------------------------------------------------------------------------------------------
\title{Compact Gaussian quantum computation by multi-pixel homodyne detection}

\author{G. Ferrini$^1$, J.P. Gazeau$^2$, T. Coudreau$^3$, C. Fabre$^1$, N. Treps$^1$}
\address{$^1$ Laboratoire Kastler Brossel, Universit\'e Pierre et Marie Curie-Paris 6, ENS, CNRS; 4 place Jussieu, 75252 Paris, France}
\address{$^2$ Laboratoire APC, Universit\'e Paris Diderot, Sorbonne Paris Cit\'e, 75205 Paris, France}
\address{$^3$ Laboratoire Mat\'eriaux et Ph\'enom\`enes Quantiques, Universit\'e Paris Diderot, Sorbonne Paris Cit\'e, CNRS, UMR 7162, 75013 Paris, France}
\ead{giulia.ferrini@spectro.jussieu.fr} 
\date{\today}

\begin{abstract}

We study the possibility of producing and detecting continuous variable cluster states in an optical set-up in an extremely compact fashion. 
This method is based on a multi-pixel homodyne detection system recently demonstrated experimentally, which includes classical data post-processing. 
It allows to incorporate the linear optics network, usually employed in standard experiments for the production of cluster states, in the stage of the measurement.
After giving an example of cluster state generation by this method, we further study how this procedure can be generalized to perform gaussian quantum computation.

\end{abstract}

%\pacs{still to be checked!}
\maketitle
%-------------------------------------------------------------------------------------------------------------------------------------------------------------------------
\section{Introduction}

Quantum computing is a very promising subject despite the formidable challenges it must overcome~\cite{Science_spec_issue}. 
One of those challenges lies in the difficulty of finding a system allowing for a large number of entangled states to be manipulated and for many quantum gates to be implemented successively.
In this context, an interesting avenue consists in measurement--based quantum computing~\cite{Briegel01,Nielsen06,Raussendorf01} (MBQC) protocols which rely on the availability of a large, multipartite entangled state on which a series of measurements is performed. For each operation that we wish to implement, a specific sequence of measurements on its modes has to be chosen. The measurement outcomes are in general used to determine which measurement has to be performed afterwards, and to perform a final correction stage on the output state.
Optical systems are promising candidates for the experimental realization of MBQC \cite{Menicucci06,Menicucci_PRA_09} and convincing demonstrations have already been made, both in the discrete~\cite{Walther} and continuous variable regime~\cite{Ukai11}. 
Continuous variable (CV) systems are especially promising since they allow to create and to process cluster states in a deterministic fashion, contrarily to the discrete case~\cite{V. Loock_rev}. 

Most often, the creation of multimode entangled states such as cluster states in a CV quantum optical setup is implemented using a series of squeezers followed by a network of beam-splitters and dephasers, which transform the squeezed input modes into entangled output modes~\cite{van_Loock_PRA_08, Ukai11, Aoki,Su12}. 
The configuration of this network varies considerably with the state to be generated, and its complexity grows rapidly with the number of modes, which renders this method poorly scalable. 
%Later on, the generated state is probed by local measurements on each mode, according to suitably chosen observables~\cite{Menicucci_PRA_09}. 

In a recent experiment at Australian National University~\cite{Canberra}, some of us have experimentally demonstrated the possibility of performing exceptionally compact operations on optical modes using spatial (transverse) modes of light.
In this experiment, multiple transverse modes were generated and individually squeezed in optical cavities, and then mixed together to produce a multimode squeezed beam.
This beam was then measured by means of a multi-pixel homodyne detection (MPHD) system, followed by digital post-processing of the acquired signals, \textit{e.g.} multiplication of each signal a by suitable gain~\cite{Beck}. 
This allowed them to  measure simultaneously an arbitrary quadrature on all the modes in a chosen mode set. 
The use of this method to emulate cluster states statistics was also sketched in this work.

It is also possible to generate multimode quantum states of light in the frequency domain starting from an optical femtosecond frequency comb and using a synchronously pumped optical parametric oscillator (SPOPO).
In this context, the formation of three-mode squeezed non-classical states has already been demonstrated~\cite{Pinel}. 
A multimode detection system analogous to the one employed in Ref.~\cite{Canberra} can be implemented for the detection of these frequency modes, where the spatial pixels are replaced by frequency bins.
An appealing advantage of this setup is that the source is now scalable, as all the modes are produced within only one optical device, and potentially hundreds of modes can be entangled.

This motivates our study of the ensemble of equivalent unitary operations which can be readily implemented by the multi-pixel detection followed by digital signal recombination. 
In particular, we give a procedure to assess whether and how a given cluster state can be realized with this method. Then, we apply this idea to perform gaussian quantum computation in the measurement-based model and give some specific examples.
%-------------------------------------------------------------------------------------------------------------------------------------------------------------------------
\section{Modelization of the operations which can be performed on the modes}

\begin{center}
\begin{figure}[h!]
a)
\begin{minipage}{\columnwidth}
\centering
\includegraphics*[width=0.65\columnwidth]{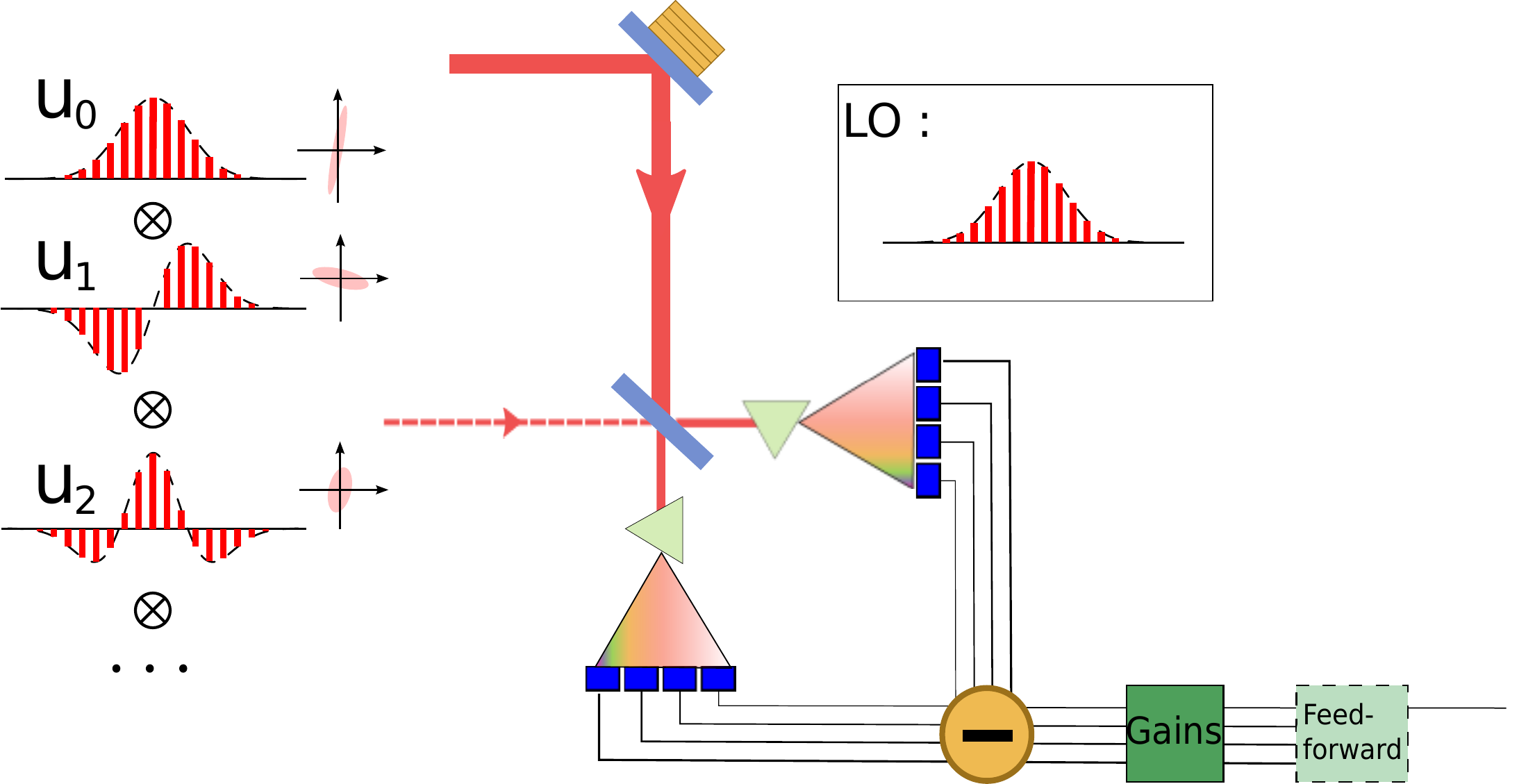}
\end{minipage}
\begin{minipage}{\columnwidth}
\centering
\vspace{0.3cm}
{\huge $\Updownarrow$}
\end{minipage}
\vspace{-0.8cm}
b)
\begin{minipage}{\columnwidth}
\centering
\includegraphics*[width=\columnwidth]{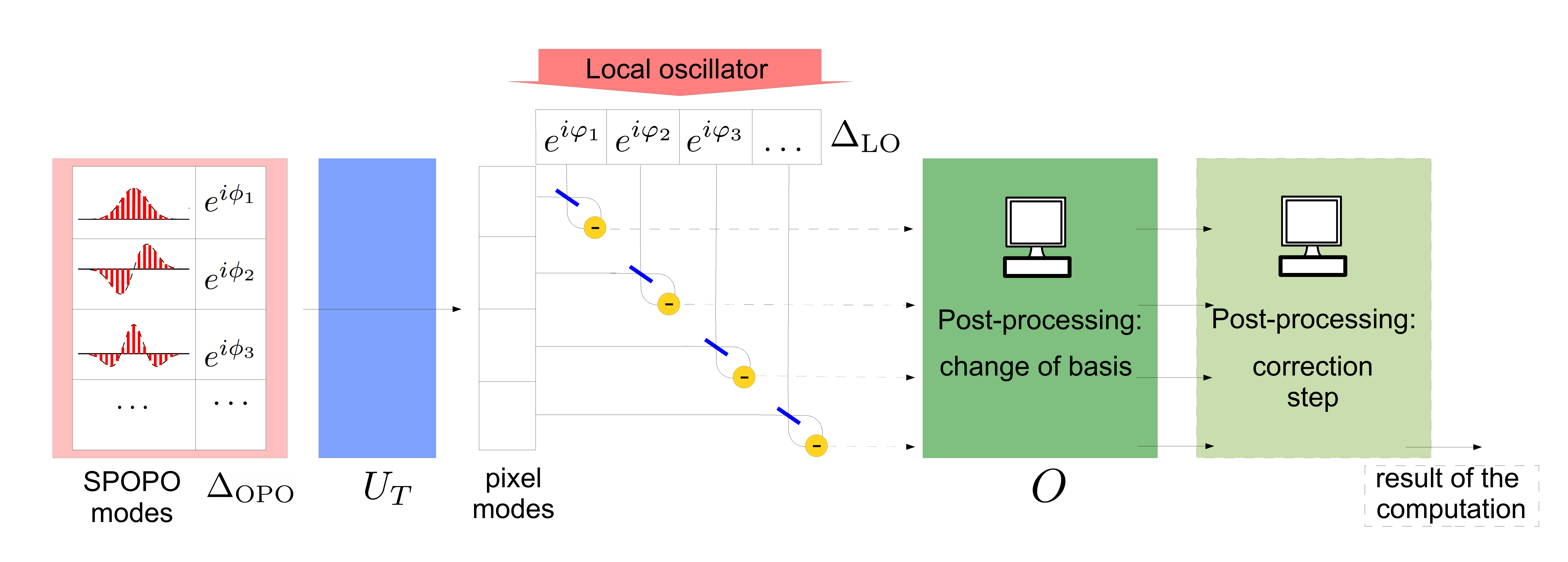}
\end{minipage}
\vspace{0.8cm}
\caption{a) Sketch of the experiment performed with frequency modes. The (infinitely) $\hat p$-squeezed modes $u_i(\rho)$ generated by synchronously pumped optical parametric oscillator (SPOPO) are mixed with a local oscillator on a 50/50 beamsplitter. Both outputs are then dispersed and each frequency band is sent to a different pixel of the two multi-pixel detectors. The differences of the acquired signals are taken pixel by pixel. These are later processed by an ordinary computer which multiplies every trace by a suitable gain, for instance to obtain a cluster state. The additional correction step indicated in the dashed box is only required when performing a quantum computation; it reinterprets the measurement result on the mode of interest depending on the outcomes of the measurements on the other modes. Note that an analog device can similarly be implemented for the detection of spatial input modes. b) Functional modelization of the same experiment. $\Delta_{\mathrm{OPO}}$ models squeezing quadratures of source modes, $U_T$  light propagation from source to multi pixel detector, $\Delta_{\mathrm{LO}}$ local oscillator phases in the pixel basis and $O$ digital data processing. Refer to main text for details.}
\label{fig1}
\end{figure}
\end{center}

Let us first model an experiment that would consist in generating a multimode squeezed state and in performing a multi-pixel homodyne detection from the point of view of canonical transformations on the optical input modes. 
Such an experiment can be separated in four steps, as can be seen in Fig.\ref{fig1}: a) generating a composite beam made of independent orthogonal squeezed modes, which can be either spatial or frequency modes; b) mixing it on a beam splitter with a shaped local oscillator; c) propagating it to multi-pixel detectors; d) recombining digitally the acquired signals.

As discussed above, there are several ways of generating multimode squeezed beams. Nevertheless, independently of the choice adopted, the result can be modeled as an optical field composed of $N$ orthogonal and normalized modes of the electro-magnetic field $u_i(\rho)$, where $\rho$ could either be spatial coordinates $(x,y)$ or frequency components $\omega$ depending on the realization. We assume that in each mode the state of the field is infinitely squeezed, say along the $\hat p$ quadrature, corresponding hence to the $\hat p$-eigenstate $|  s \rangle_p$, being $\hat p |  s \rangle_p = s |  s \rangle_p$, with zero eigenvalue, i.e. $|  0 \rangle_p = 1/\sqrt{2 \pi} \int ds  | s\rangle_q$~\cite{Braunstein_paper}.
The discussion of the effect of finite squeezing is beyond the scope of this paper.

These $N$ modes are accompanied by vacuum modes in all the other modes of the complete modal basis $\{u_i(\rho)\} $. To each mode are associated a pair of bosonic operators $\hat a_{u_i}$ and $\hat a^\dagger_{u_i}$ with %$a_{u_i} = \int \hspace{-0.2cm} \int_{\rho \in \R^2} d^2 \rho \hat{a}(\rho) u^{*}_{i}  (\rho)$. 
$\hat{a}_{u_i} = \int  d \rho \hat{a}(\rho) u^{*}_{i}  (\rho)$ (where $\int  d \rho$ stands for $\int \hspace{-0.2cm} \int_{\rho \in \R^2} d^2 \rho$ if $\rho$ is a $2D$ variable), and $\hat{a}(\rho) = \sum_{k=1}^{\infty}  \hat{a}_{u_k} u_k  (\rho) $.
These input modes can be grouped in a vector ${\bf \hat a_{u}} = (\vec{a}_{u}, \vec{a}_{u}^{\dagger})^T = (\hat{a}_{u_1},\hat{a}_{u_2}, ... ,\hat{a}_{u_1}^{\dagger}, \hat{a}_{u_2}^{\dagger}...)^T$. 

Once the multimode squeezed beam is generated, it is mixed on the beam splitter with a local oscillator. Then, it is propagated to two arrays of detectors composed of pixels, each of surface $S_i$, that collect the intensity of the incoming light from the two respective beams, and the differences pixel by pixel of the two signals are taken. This propagation is governed by diffraction in the spatial case, and by frequency dispersion realized with the help of a prism in the frequency case (see Fig.\ref{fig1}). 
To describe the effect of the detection of the beam it is convenient to introduce another set of modes, that we shall call {\it pixel modes}; these are defined as 
\be
\label{eq:pixel_modes}
v_{i} (\rho)= 
\left\{
\begin{array}{rl}
\kappa_i \, \,  u_{\mathrm{LO}}   (\rho) &  \hspace{0.4cm} \forall \rho \in S_{i} \\
 0       &                                           \mbox{    elsewhere}.
\end{array}
\right. ,
\ee
%where $\chi_{S_i}(\rho) = 1$ for $\rho \in S_i$ and $0$ otherwise (i.e., $\chi_{S_i}$ is the window function associated to pixel $i$), 
and correspondingly $\hat{a}_{v_i} = \int d \rho \hat{a}(\rho) v_{i}^*  (\rho)$, where $\kappa_i$ is a normalizing constant such that $  \int_{\rho \in \R} d \rho v_{i}^2 (\rho) =  \kappa_i^2 \int \hspace{-0.2cm} \int_{\rho\in S_{i}} d^2  \rho \, | u_{\mathrm{LO}} (\rho) |^2 = 1$. Eq.(\ref{eq:pixel_modes}) implies that each pixel mode is a "slice" of the local oscillator, i.e. it shares its shape but only on a window with width given by the pixel dimension (and it is zero ouside).
Note that the pixel modes form a complete basis, allowing to reconstruct all the possible modes, only in the limit of infinite pixel number.
%with $v_i (\rho) = \kappa u_{\mathrm{LO}}(\rho) \chi_{S_i} (\rho)$, 
The detection process can be modeled as an effective transformation corresponding to light propagation from the squeezed modes set to the pixel modes one, and to a homodyning in that basis with an intense local oscillator $u_{\mathrm{LO},1}(\rho)$. Here for consistency we assume that $\rho$ is, from the beginning, the variable that spatially describes the field in the multi-pixel detector plane. For instance, in Fig. \ref{fig1} it does indeed correspond to frequency $\omega$, but in the spatial case it depends on the actual imaging system between squeezing generation and the detectors.
This transformation is described by a matrix $U_T$ acting on the bosonic operators: $\vec{a}_{v} = U_T \vec{a}_{u}$, which brings the input modes $\hat{a}_{u_i}$  onto the pixel modes $\hat{a}_{v_i}$, with ${U_T}_{ij} =  \kappa _i \int_{\rho \in S_i} d \rho \,  u^*_{\mathrm{LO}}(\rho) u_j(\rho)$. More details can be found in \ref{sse:mhd}, and in Refs.~\cite{Jeff_thesis,Canberra}. Note that in principle the detection matrix ${U_T}$ does not need to be square, since the number of pixels $P$ may differ from the number of input modes - the latter may be potentially infinite. 

\begin{center}
\begin{figure} [h!]
\begin{minipage}{.24\columnwidth}
\includegraphics*[width=\columnwidth]{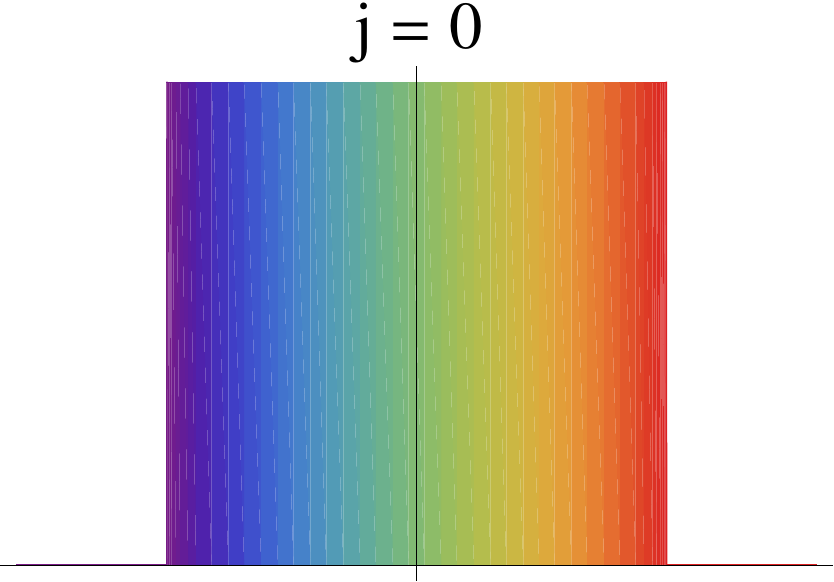}
\end{minipage}
\begin{minipage}{.24\columnwidth}
\includegraphics*[width=\columnwidth]{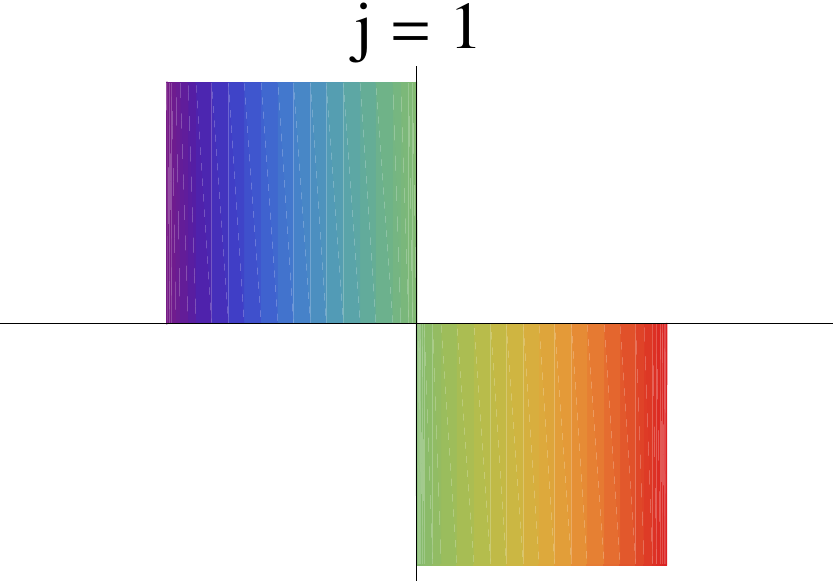}
\end{minipage}
\begin{minipage}{.24\columnwidth}
\includegraphics*[width=\columnwidth]{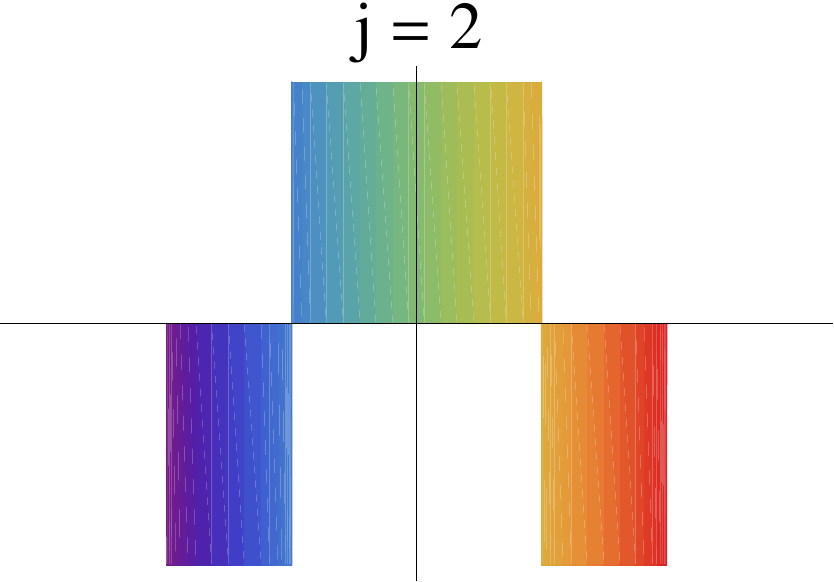}
\end{minipage}
\begin{minipage}{.24\columnwidth}
\includegraphics*[width=\columnwidth]{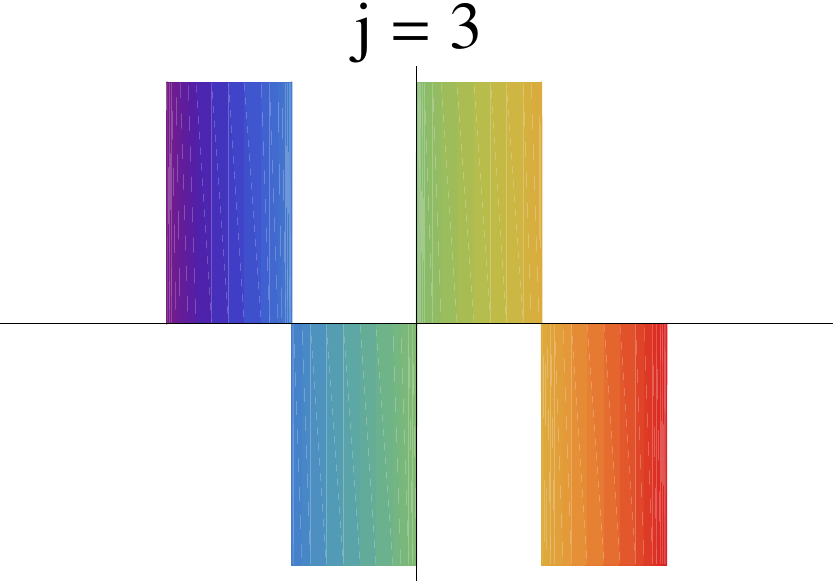}
\end{minipage}\\
\vspace{0.5cm}
\begin{minipage}{.24\columnwidth}
\includegraphics*[width=\columnwidth]{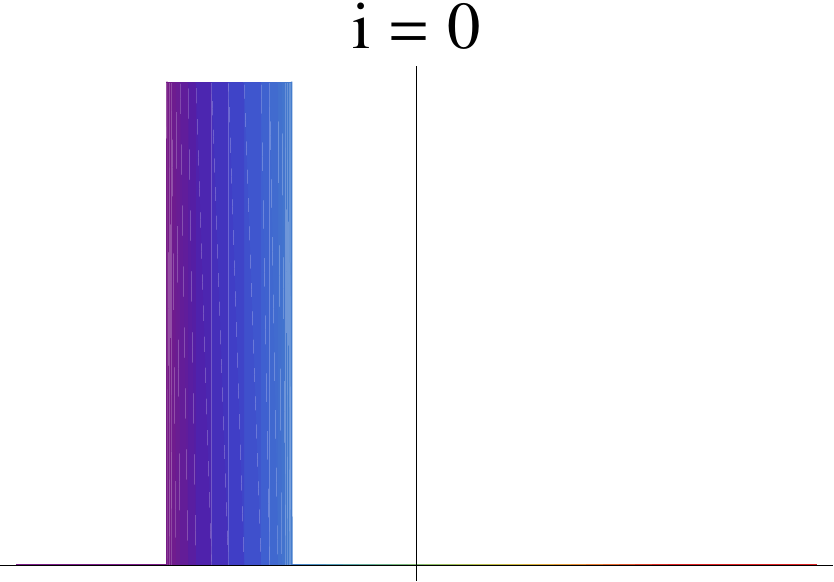}
\end{minipage}  
\begin{minipage}{.24\columnwidth}
\includegraphics*[width=\columnwidth]{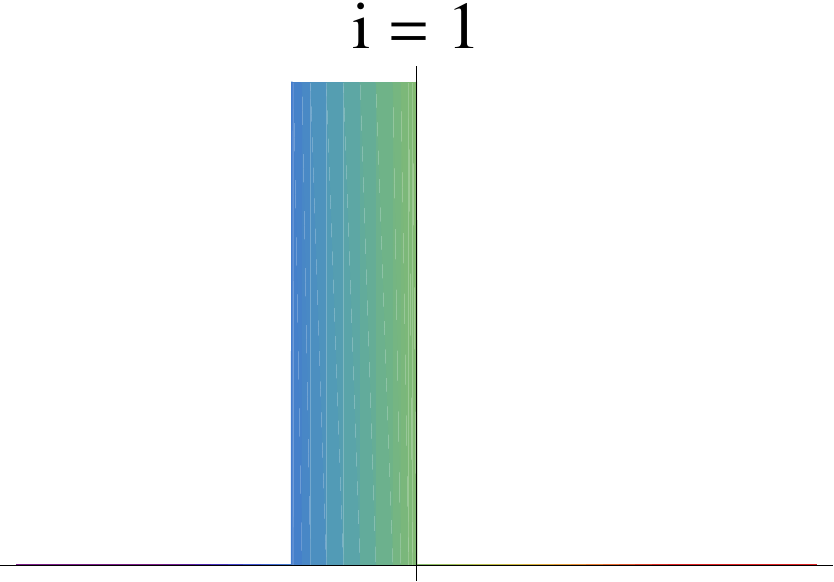}
\end{minipage}
\begin{minipage}{.24\columnwidth}
\includegraphics*[width=\columnwidth]{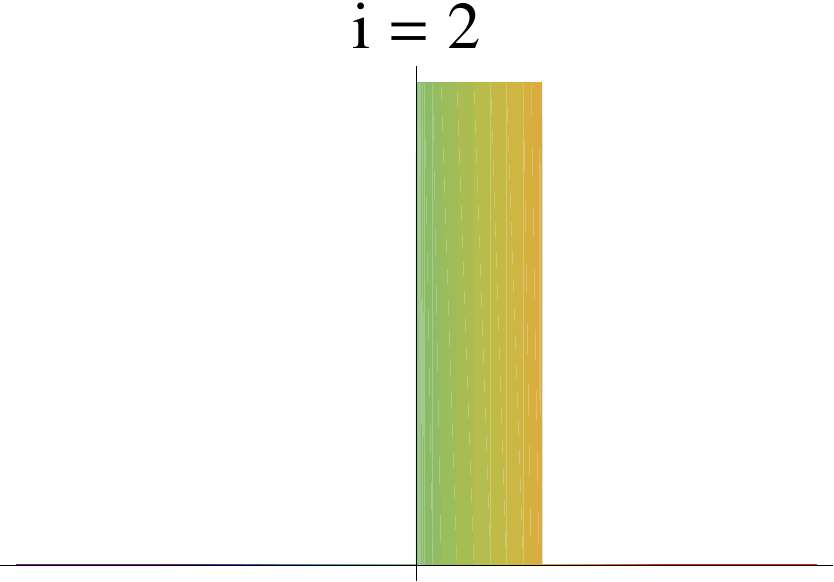}
\end{minipage}
\begin{minipage}{.24\columnwidth}
\includegraphics*[width=\columnwidth]{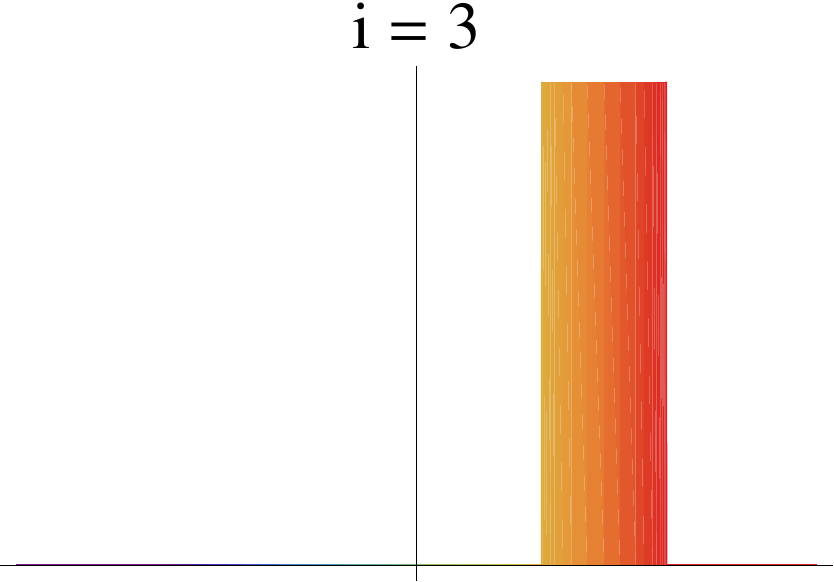}
\end{minipage}
\caption{Input square flip-modes $u_j (\rho)$ (top panels) and pixel modes $v_i (\rho) = \sum_i {U_T}_{ij} u_j (\rho)$ (bottom panels) for $0 \leq i,j \leq 3$, as a function of $\rho$. As it can be seen, the latter allow to reconstruct the profile of the former and conversely.}
\label{fig2}
\end{figure}
\end{center}

As an example to fix the ideas, that we will use later, consider the simple case in which four input squeezed modes are detected with the help of four pixel modes. Let us assume that the intensity distribution of the local oscillator mode is the first mode of this basis, i.e. $u_{\mathrm{LO}}(\rho) = u_{1}(\rho)$. We take a toy model for the input modes, describing the mode $n$ by a square profile in which we introduce $n -1$ $\pi$-phase shifts (or ``flips") at regular intervals, as illustrated in Fig.\ref{fig2} (top panels). These modes are however qualitatively similar to the modes generated by a SPOPO, which in turns are close to Hermite-Gauss modes~\cite{Pinel}.
%, leading to the pixel modes sketched in Fig.\ref{fig1} (bottom panels).
%In this case, the pixel basis is complete for the input mode basis (truncated at $3 \sigma$), and the detection process is then described by a finite-dimensional square matrix which reads (in its normalized version)
In this simple case, the matrix  $U_T$ is unitary and takes the simple form 
\ba
\label{eq:Am1_N4again}
U_T = \fr{1}{2}
\left( 
\begin{array}{cccccccc}
1 & 1 & -1 & 1 \\
1 & 1 &  1 & -1 \\
1 & -1 & 1 & 1 \\
1 & -1 & -1 & -1 \\
\end{array}
\right). 
\ea 

 %\\$\spadesuit$ {\bf Can I say unitary if it is not square? How can the pixel basis be complete if the input basis is infinite? In which case do I need a infinite basis?}
 
Above we have assumed that the input state is $\hat p$-squeezed on each mode. Experimentally it is in principle possible to modify the squeezing quadrature. For simplicity, we model that by changing the input mode phase instead of its state, which is completely equivalent in an homodyne detection scheme. This is done by introducing a diagonal dephasing matrix $\Delta_{\mathrm{OPO}}$ with ${\Delta_{\mathrm{OPO}}}_{i,j} = e^{i \phi_i} \delta_{ij}$. The input modes entering the detection system are $u_i'(\rho) = e^{-i \phi_i} u_i(\rho)$ and $\vec{a}_{u'} = e^{i \phi_i}   \vec{a}_{u}$. The transformation of these input modes to the pixel set as defined above is now $\vec{a}_{v} = U_T \Delta_{\mathrm{OPO}}^*  \vec{a}_{u'}$, as $\Delta_{\mathrm{OPO}}^*$ moves the phase shifted modes to the modes $\vec{a}_{u}$ and $U_T$ is the transformation from $\vec{a}_{u}$ to $\vec{a}_{v}$.

With the same approach, the ability of shaping the phase of the local oscillator by phase-shifting the pixel modes can be modeled by the action of a diagonal matrix $\vec{a}_{v'} = \Delta_{\mathrm{LO}} \vec{a}_{v}$ with elements ${\Delta_{\mathrm{LO}}}_{i,j} =  e^{i \varphi_i} \delta_{i,j}$, leading to the modes ${v'_{i}} (\rho)= e^{-i \varphi_i} {v_{i}}(\rho)$.

Finally, after the detection, it is possible to digitally recombine the electronic signals coming from each pixel by multiplying them with real gains~\cite{Beck}. 
This allows us to detect the field according to desired modes (e.g., as we shall see later, according to the modes which correspond to a cluster state), and amounts to applying a $P$-dimensional orthogonal matrix, i.e. $\vec{a}_{v^{''}}  =  O \vec{a}_{v'} \equiv \vec{a}_{\mathrm{out}} $ (see also the Appendix).

In summary, we find that the series of the possible transformations that we can perform on the input modes are 
\be
\label{eq:trtot_m}
\vec{a}_{\mathrm{out}} =  O \Delta_{\mathrm{LO}} G \, \, \vec{a}_{u} \equiv U_{\mathrm{MPHD}} \, \,  \vec{a}_{u},
\ee
where $\Delta_{\mathrm{LO}} = \mathrm{diag}(e^{i\varphi_1}\, ,\, e^{i\varphi_2}\, , \,\ldots\, e^{i\varphi_N})$, % $\sum_{i=1}^{N}\varphi_i = 0$
$O$ is real orthogonal, %with unit determinant, i.e. $O \in SO(N)$
and $G = U_T \Delta_{\mathrm{OPO}}^*$ is a fixed matrix.
%In what follows, we are going to consider the case in which the number of pixels equals the (finite) number of input modes, $P = N'$ (we recall that $N$ input modes are squeezed). An explicit example will be provided below.
Note that since $ \Delta_{\mathrm{LO}}  \Delta_{\mathrm{LO}}^\dagger = \mathcal  I$ and $G G^{\dagger} = \mathcal I $ the product $  O \Delta_{\mathrm{LO}} G = U_{\mathrm{MPHD}}$ is unitary. As implicit in the previous discussion, the matrices $O$ and $ \Delta_{\mathrm{LO}}$ are easily tunable in the experiments.  The dephasing matrix $\Delta_{\mathrm{OPO}} $ can also be adjusted to some extent by changing the spectral shape of the pump~\cite{Giuseppe}. 

%-------------------------------------------------------------------------------------------------------------------------------------------------------------------------
\section{Characterization of the feasible transformations}

We now address the characterization of the sub-set of unitary operations, acting on the annihilation operator $\vec{a}_{u}$, which can be emulated by means of Eq.(\ref{eq:trtot_m}). 
The question: ``which class of unitary transformations can we emulate in the laboratory by means of the multi-pixel homodyne detection plus data processing?" can be recast into the following question: ``given a unitary matrix $U_{\mathrm{th}}$, can it be written in the form of Eq.(\ref{eq:trtot_m})?" Answering this questions requires finding a solution for the parameters of $O$ and $\Delta_{\mathrm{LO}}$ (given a certain fixed detection matrix $G$) such that 
\be
\label{eq:trasf_labo_bis}
U_{\mathrm{th}} = U_{\mathrm{MPHD}}.
\ee
It is easily seen that a necessary and sufficient condition for Eq.(\ref{eq:trasf_labo_bis}) is
\be
\label{eq:cod}
{U_{\mathrm{th}}'}^T U_{\mathrm{th}}' = D,
\ee
with $U_{\mathrm{th}}' = U_{\mathrm{th}} G^{\dagger}$ and where $D$ is a diagonal matrix with unit modulus complex elements. 
The necessity is simply proved by assuming that Eq.(\ref{eq:trasf_labo_bis}) holds, and by direct calculation of the product ${U_{\mathrm{th}}'}^T U_{\mathrm{th}}'$ upon substitution of Eq.(\ref{eq:trtot_m}), which yields ${U_{\mathrm{th}}'}^T U_{\mathrm{th}}' = G^* G^T \Delta_{\mathrm{LO}}^T O^T O \Delta_{\mathrm{LO}} G G^{\dagger} = \Delta_{\mathrm{LO}}^2$, where we have used $\Delta_{\mathrm{LO}}^T = \Delta_{\mathrm{LO}}$ since this matrix is diagonal, and where $\Delta_{\mathrm{LO}}^2$ is obviously a diagonal matrix with unit modulus complex elements.
For the sufficiency, we can show that if Eq.(\ref{eq:cod}) holds, we can always find at least a set of ``experimental" matrices $\Delta_{\mathrm{LO}}$ and $O$ such that their combination of the form of $U_{\mathrm{MPHD}}$ in Eq.(\ref{eq:trtot_m}) satisfies Eq.(\ref{eq:trasf_labo_bis}). This is achieved by taking as a realization of the diagonal matrix $\Delta_{\mathrm{LO}}$ any of the solutions of the equation 
\be
\label{eq:cod1}
\Delta_{\mathrm{LO}} = D^{\fr{1}{2}} = ({U_{\mathrm{th}}'}^T U_{\mathrm{th}}')^{\fr{1}{2}}.
\ee
%as can be readily verified by substitution of the ansatz $U_{\mathrm{th}} = O \Delta$ in Eq.(\ref{eq:cod}). 
%Note that if the condition (\ref{eq:cod}) is satisfied, 
Then, it is automatically ensured that, for each solution $\Delta_{\mathrm{LO}}$, the matrix
\be
\label{eq:cod2}
O = U_{\mathrm{th}}' \Delta_{\mathrm{LO}}^{-1}  %= U_{\mathrm{th}} (U_{\mathrm{th}}^T U_{\mathrm{th}})^{-\fr{1}{2}}
\ee
is orthogonal and fulfills Eq.(\ref{eq:trasf_labo_bis}). Indeed by direct calculation we obtain $U_{\mathrm{MPHD}} = O \Delta_{\mathrm{LO}} G =  U_{\mathrm{th}}' \Delta_{\mathrm{LO}}^{-1} \Delta_{\mathrm{LO}} G =  U_{\mathrm{th}}$, which completes our proof.

As a major result of our work this shows that, given a certain unitary matrix $U_{\mathrm{th}}$, if a solution of Eq.(\ref{eq:cod}) exists, with the use of Eqs.(\ref{eq:cod1}) and (\ref{eq:cod2}) we can determine the suitable experimental parameters in terms of the gain coefficients $O$ and the local oscillator phase shaping $\Delta_{\mathrm{LO}}$. These allow to obtain with the multi-pixel homodyne detection measurement the statistics of the quadrature outcomes as if the matrix $U_{\mathrm{th}}$ - usually corresponding to a complex optical network - had been applied to the input squeezed modes, and the same quadratures were measured one by one.
Let us also note that our proposal is versatile, in the sense that different matrices $U_{\mathrm{th}}$ can be produced with little or no reconfiguration of the experimental set-up.
%-------------------------------------------------------------------------------------------------------------------------------------------------------------------------
\subsection{Cluster states in the multimode beam}

To provide a demonstration of our method, we consider the simple case in which four input squeezed modes are detected with the help of four pixel modes, which entails the mode transformation given in Eq.(\ref{eq:Am1_N4again}).
Let us consider as a concrete example the case in which the desired unitary transformation to be emulated is the one which transforms $N$ squeezed modes in a linear cluster state. 
\begin{center}
\begin{figure} [h!]
\centering
\begin{minipage}{.54\columnwidth}
\centering
\includegraphics*[width=\columnwidth]{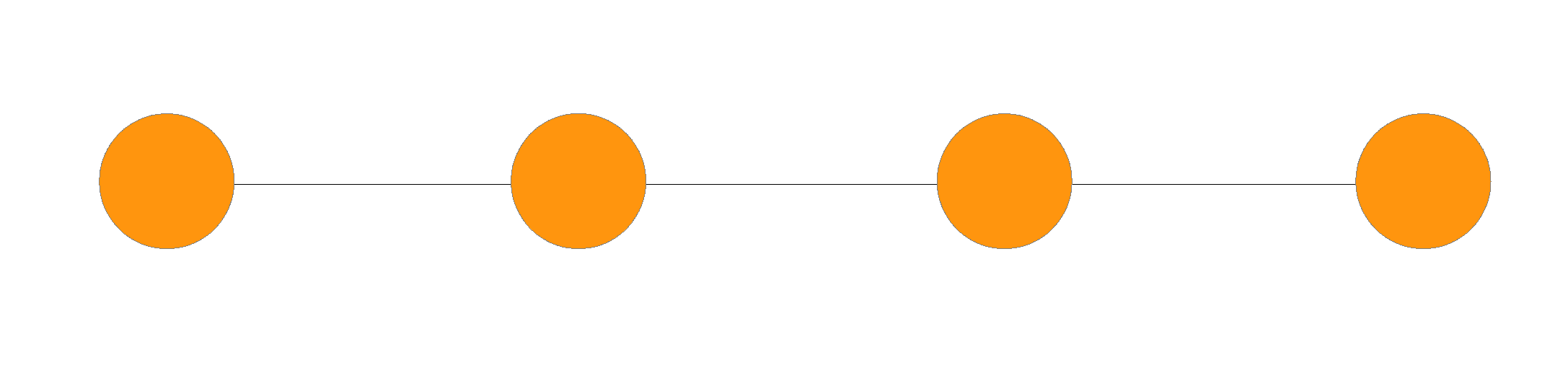}
\end{minipage}
\caption{Four-mode linear cluster state. Each circle represent a cluster mode and the edges represent entanglement between modes, namely that the two modes have been acted upon with a $C_Z$-type interaction $e^{i \hat q_i \otimes \hat q_j}$~\cite{Menicucci_PRA_09}.}
\label{fig3}
\end{figure}
\end{center}
In the case $N = 4$ (Fig. \ref{fig3}), such a transformation is represented by the following matrix~\cite{van_Loock_PRA_08}
\be
\label{eq:lin_csbis}
U_{\mathrm{th}} = U_{\mathrm{lin}} = 
\left( 
\begin{array}{cccccccc}
  \fr{1}{\sqrt{2}} & \fr{1}{\sqrt{10}} &  \fr{2 i}{\sqrt{10}}  & 0   \\
  \fr{i}{\sqrt{2}}  &  -\fr{i}{\sqrt{10}}  &  \fr{2}{\sqrt{10}}  & 0 \\
 0 &  -\fr{2}{\sqrt{10}}  &  \fr{i}{\sqrt{10}}  &  \fr{i}{\sqrt{2}}   \\
 0 &  -\fr{2i}{\sqrt{10}} &  -\fr{1}{\sqrt{10}}  &  \fr{1}{\sqrt{2}}  \\
\end{array}
\right).
\ee
%\footnote{If the detection matrix $\tilde{U}_T$ is not normalized, we can first solve the problem for its normalized version ${U}_T$ as shown in the main text, finding a solution for $O$ and $\Delta_{\mathrm{LO}}$. Then, $U_{\mathrm{th}} = O \Delta_{\mathrm{LO}} U_T \Delta_{\mathrm{OPO}} = c O \Delta_{\mathrm{LO}} \tilde{U}_T \Delta_{\mathrm{OPO}} =  O' \Delta_{\mathrm{LO}} \tilde{U}_T \Delta_{\mathrm{OPO}}$ where now the matrix to be applied by means of the electronic gains $O'$ is not orthogonal.}
In order to find a simple algebraic solution for the problem we consider the dephasing matrix $\Delta_{\mathrm{OPO}}^* = \mathrm{diag} (1, i, i, 1)$. With the use of Eqs.(\ref{eq:cod}), (\ref{eq:cod1}), (\ref{eq:lin_csbis}) and (\ref{eq:Am1_N4again}) we find
\ba
\label{eq:D_sol_linear_cluster}
{U_{\mathrm{lin}}'}^T U_{\mathrm{lin}}' &=&
 \left(
\begin{array}{cccc}
\fr{-2 -  i}{\sqrt{5}}  & 0 & 0 & 0 \\
 0 &\fr{2 -  i}{\sqrt{5}} & 0 & 0 \\
 0 & 0 & \fr{2 +  i}{\sqrt{5}} & 0 \\
 0 & 0 & 0 &  \fr{-2 +  i}{\sqrt{5}}
\end{array}
\right) .
\ea
Since the matrix in Eq.(\ref{eq:D_sol_linear_cluster}) is diagonal, the condition in Eq.(\ref{eq:cod}) is satisfied, which implies that the corresponding cluster state matrix (\ref{eq:lin_csbis}) can be implemented experimentally. 
Among the $2^N = 16$ possible solutions for ${\Delta_{\mathrm{LO}}}_{\mathrm{lin}}$ and the corresponding orthogonal matrix $O_{\mathrm{lin}}$ satisfying  Eqs.(\ref{eq:cod1}, \ref{eq:cod2}) for $U_{\mathrm{th}} = U_{\mathrm{lin}}$ we can chose for instance the set
 {\footnotesize 
\ba 
\hspace{-2.5cm}
  {{\Delta_{\mathrm{LO}}}_{\mathrm{lin}}}_1 &=& 
\left(
\begin{array}{cccc}
 e^{i\fr{\gamma + \pi}{2}} & 0 & 0 & 0 \\
 0 &  e^{-i\fr{\gamma}{2}} & 0 & 0 \\
 0 & 0 & e^{i\fr{\gamma}{2}} & 0 \\
 0 & 0 & 0 & e^{-i\fr{\gamma + \pi}{2}}
\end{array}
\right) = 
\left(
\begin{array}{cccc}
 -0.23+0.97 i & 0 & 0 & 0 \\
 0 & 0.97-0.23 i & 0 & 0 \\
 0 & 0 & 0.97+0.23 i & 0 \\
 0 & 0 & 0 & -0.23-0.97 i
\end{array}
\right); \nn \\
\hspace{-2.5cm}
{O_{\mathrm{lin}}}_1 &=& \fr{1}{\sqrt{2}}
\left(
\begin{array}{cccc}
 -\cos \alpha & \sin \alpha & \sin \alpha & -\cos \alpha \\
\sin \alpha & -\cos \alpha & \cos \alpha & -\sin \alpha \\
\sin \alpha & \cos \alpha & \cos \alpha & \sin \alpha \\
 -\cos \alpha & -\sin \alpha & \sin \alpha & \cos \alpha
\end{array}
\right) = 
\left(
\begin{array}{cccc}
 -0.16 & 0.69 & 0.69 & -0.16 \\
 0.69 & -0.16 & 0.16 & -0.69 \\
 0.69 & 0.16 & 0.16 & 0.69 \\
 -0.16 & -0.69 & 0.69 & 0.16
\end{array}
\right)
%\frac{1}{2} \left(
%\begin{array}{cccc}
 %- \sqrt{1-\frac{2}{\sqrt{5}}} &  \sqrt{1+\frac{2}{\sqrt{5}}} & \sqrt{1+\frac{2}{ \sqrt{5}}} & - \sqrt{1-\frac{2}{\sqrt{5}}} \\
 %\sqrt{1+\frac{2}{\sqrt{5}}} & - \sqrt{1-\frac{2}{\sqrt{5}}} & \sqrt{1-\frac{2}{\sqrt{5}}} & -   \sqrt{1+\frac{2}{\sqrt{5}}} \\
 %\sqrt{1+\frac{2}{ \sqrt{5}}} &  \sqrt{1-\frac{2}{\sqrt{5}}} & \sqrt{1-\frac{2}{\sqrt{5}}} & \sqrt{1+\frac{2}{\sqrt{5}}} \\
 %- \sqrt{1-\frac{2}{\sqrt{5}}} & - \sqrt{1+\frac{2}{\sqrt{5}}} & \sqrt{1+\frac{2}{ \sqrt{5}}} & \sqrt{1-\frac{2}{\sqrt{5}}}
%\end{array}
%\right)} \nn
 \ea } 
with $\gamma \equiv \arctan{\fr{1}{2}}$ and $\alpha = \cos ^{-1} \sqrt{\frac{1}{2} \left(1-\frac{2}{\sqrt{5}}\right)} $. Hence, we conclude that shaping the phase of the local oscillator on each pixel modes as prescribed by the matrix $ {{\Delta_{\mathrm{LO}}}_{\mathrm{lin}}}_1$ in Eq.(\ref{eq:Am1_N4again}) and choosing the digital recombination gains according to ${O_{\mathrm{lin}}}_1$ allows us to obtain a statistics for the quadratures measurement as if a network of beam-splitters generating a linear cluster state had been applied to the input squeezed modes.
We have checked that a numerical solution exists for other choices of the dephasing matrix, namely for the one corresponding to the physical dephasings of the cavity modes in the experiment of Ref.~\cite{Pinel}.

%-------------------------------------------------------------------------------------------------------------------------------------------------------------------------
\section{Simple examples of quantum computations}

We now address the emulation of a simple example of quantum computation in the measurement-based model (MBQC). Since we are performing homodyne detection on all the modes, we are restricted to gaussian operations ~\cite{Menicucci_PRA_09}. This is a particularly simple framework to start with, because the implementation of gaussian operations in the MBQC model does not require an effective adaptivity - usually necessary to implement operations deterministically - and the measurements can be performed on all the modes simultaneously. The outcomes are recorded, and the ``adaptation" of the measurement basis depending on the outcomes of previous measurement can be recovered in a further classical processing step (e.g., by electronically correcting  the photocurrents, see dashed box in Fig.\ref{fig1})~\cite{Menicucci_PRA_09}. 

\begin{center}
\begin{figure} [h!]
\centering
\begin{minipage}{.54\columnwidth}
\centering
\includegraphics*[width=\columnwidth]{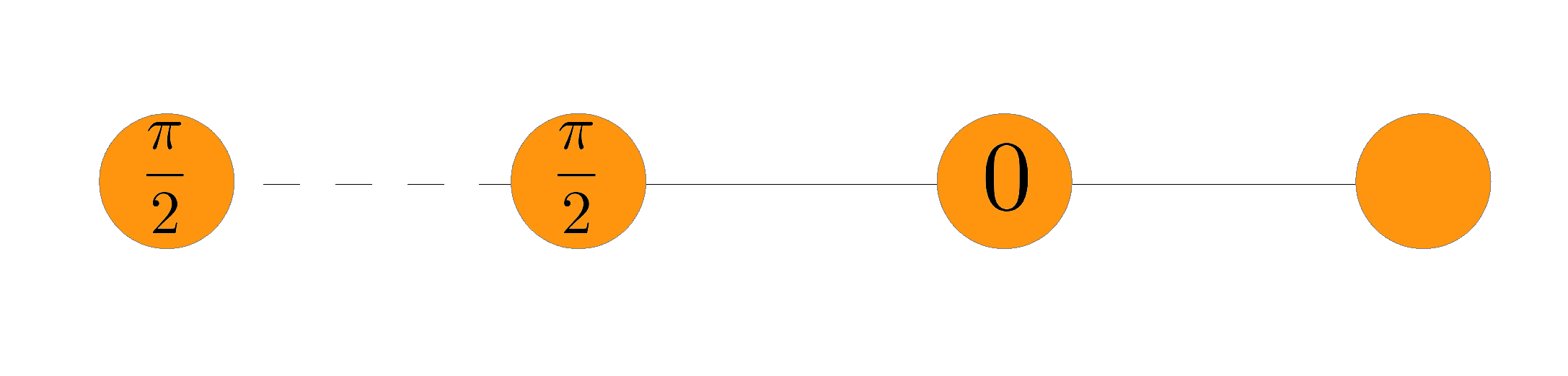}
\end{minipage}
\caption{Fourier transform of an input state through measurement-based quantum computation. The input mode is coupled via a beam-splitter (dashed line) to the first mode of a three-mode cluster state, and then these two modes are simultaneously measured, which teleport the input state onto the second mode. Then, a suitable quadrature measurement is performed on the second mode of the cluster, projecting the last mode onto the desired output. With the notation used in this picture~\cite{Bruss_book}, the angles appearing on each mode specify which quadrature $\hat{q'}_{i} \sin \theta_{i} + \hat{p'}_{i} \cos \theta_{i}$ shall be measured on it in order to implement the desired operation.}
\label{fig4}
\end{figure}
\end{center}
We consider here single-mode operations.  A quantum computation implies first coupling an initial quantum state to be processed to a cluster state, and then suitable measurements of the quadratures on each mode of the cluster (Fig.\ref{fig4}). The measurement of the last mode is not part of the computation but provides a read-out of the output state, the ``result" of the computation. The coupling between the initial state and the first mode of the cluster can be realized by a generalized teleportation~\cite{Furusawa_Loock_th}: these two modes are coupled by a beam-splitter interaction, eg. $\left(\begin{array}{cccccccc}
\hat{a}_{in}'  \\
\hat{a}_{1}' \\
\end{array}
\right) 
= U^{in,1}_{\mathrm{BS}}
\left(\begin{array}{cccccccc}
\hat{a}_{in}  \\
\hat{a}_{1} \\
\end{array}
\right)$ with $U^{in,1}_{\mathrm{BS}} =  \fr{1}{\sqrt{2}}
\left(\begin{array}{cccccccc}
1 & i  \\
i  &  1 \\
\end{array}
\right) $. Measuring the quadratures $\hat{q'}_{in} \sin \theta_{in} + \hat{p'}_{in} \cos \theta_{in}$ and $\hat{q'}_{1} \sin \theta_{1} + \hat{p'}_{1} \cos \theta_{1}$ allows to transfer the input state on the second mode of the cluster state modulo a transformation $M_{tele}(\theta_{in},\theta_{1})$ controllable via the choice of the measurement angles $\theta_{in}, \theta_{1}$~\cite{Furusawa_Loock_th}\footnote{The teleported state will be actually display a quadrature shift according to some by-product operator which depends on the outcomes of the quadrature measurements on modes $in$, $1$. This can be corrected with classical post-processing at the end of the computation.}. This corresponds to the effective transformation of the quadratures  
\be
\label{eq:tele_matrix0}
\hspace{-2cm}
\left( 
\begin{array}{cccccccc}
\hat{q}^{'} \\
\hat{p}^{'} \\
\end{array} 
\right) =
M_{tele}(\theta_{in},\theta_{1}) 
\left( 
\begin{array}{cccccccc}
\hat{q} \\
\hat{p} \\
\end{array} 
\right)
\mbox{           with         }
M_{tele}(\theta_{in},\theta_{1}) =
- \left(\begin{array}{cccccccc}
\fr{\cos \theta_+}{\cos \theta_-} & \fr{\sin \theta_- - \sin \theta_+}{\cos \theta_-}  \\
\fr{\sin \theta_- + \sin \theta_+}{\cos \theta_-} &  \fr{\cos \theta_+}{\cos \theta_-} \\
\end{array}
\right)
\ee
and $\theta_{\pm} = \theta_{in} \pm \theta_1$ (the difference with Eq.(8) of Ref.~\cite{Furusawa_Loock_th} being due to a different definition of the angles $\theta_{in}, \theta_1$).
In particular, the choice $\theta_{in} = \theta_{1} = \pi/2$ teleports exactly the input state onto the second mode of the cluster, i.e. 
\be
M_{tele} \lt \theta_{in} = \fr{\pi}{2},\theta_1 = \fr{\pi}{2} \rt= 
\left(\begin{array}{cccccccc}
1 & 0 \\
0 & 1 \\
\end{array}
\right).
\ee

Let us consider as an example of quantum computation the Fourier transform of the initial state. Together with the quadrature displacement $D_{1,q}(s) = e^{i s \hat q}$ and the shear $D_{2,q}(s) = e^{i s \hat q^2}$ this is one of the elementary operations for universal one-mode gaussian quantum computation~\cite{Menicucci_PRA_09}, and it has been recently implemented experimentally with the use of a beam splitter network generating a four-mode cluster state ~\cite{Ukai11}. The Fourier transform that we wish to implement should act on the quadratures of the input state $\left( 
\begin{array}{cccccccc}
\hat{q} \\
\hat{p} \\
\end{array} 
\right)$ by rotating them, i.e. $F = \left(\begin{array}{cccccccc}
0 & -1 \\
1 & 0 \\
\end{array}
\right)$.
For a general cluster state it has been shown~\cite{Menicucci_PRA_09,Furusawa_Loock_th} that measuring the quadrature \be
\label{eq:measure_quadr_phase_gate}
%\hat{p}_{s  \hat{q}^2} = e^{-i s  \hat{q}^2} \hat p e^{i s  \hat{q}^2} 
D^{\dagger}_{q,2}(s) \hat p D_{q,2}(s) = \hat p + s \hat q = g (\hat x \sin \theta + \hat p \cos \theta)
\ee
on one of its modes, with $g = \sqrt{1 + s^2}$ and $\theta = \arctan s$, effectively transfers to the adjacent (say, right hand) mode a state transformed according to the matrix
\be
\label{eq:m_i}
M(s) =
\left(\begin{array}{cccccccc}
-s & -1 \\
1 & 0 \\
\end{array}
\right),
\ee
apart from ``by-product" displacements which are not accounted for by Eq.(\ref{eq:m_i}). These may be corrected in the end of the computation, although this is not strictly necessary~\cite{Menicucci_PRA_09}. Note that the measurement of the quadrature in Eq.(\ref{eq:measure_quadr_phase_gate}) corresponds to the measurement of the quadrature $\hat p$ after the transformation $\hat a \rightarrow e^{i \theta} \hat a$ (modulo the scale factor $g$).

We see that in the simple case under consideration a single computational measurement, to be performed on the mode $2$ of the cluster state in Fig.~\ref{fig4}, is necessary after the teleportation step. Indeed, with $s_2 = 0$ we obtain from Eq.(\ref{eq:m_i})
\be
\label{eq:sol_tf_loock}
M(s_2 = 0) = 
\left(\begin{array}{cccccccc}
0 & -1 \\
1 & 0 \\
\end{array}
\right),
\ee 
leading to $M(s_2 = 0) M_{tele}(\theta_{in} = \fr{\pi}{2},\theta_1 = \fr{\pi}{2}) = F$. 
As expressed by Eq.(\ref{eq:measure_quadr_phase_gate}), this corresponds to the measurement of the quadrature $\hat p$ on the mode $2$. As well as for the measurements implementing the teleportation, the corresponding outcome should be recorded to implement the final correction step above mentioned~\cite{Menicucci_PRA_09,Furusawa_Loock_th}. 

Hence, to summarize, in order to implement a Fourier transform on an initial state, we have to couple it via a beam-splitter interaction on a three-mode cluster state, and then to measure on each mode the quadrature specified by the angles $(\theta_{in},\theta_1,\theta_2,\theta_3) = (\fr{\pi}{2},\fr{\pi}{2},0,\theta_3)$ respectively (see Fig.~\ref{fig4}). The measurement on the third mode of the cluster yields an outcome probing the implementation of the desired transformation, and hence it can be performed according to any quadrature.

As anticipated, we take the perspective of measuring (simultaneously) the quadrature $\hat p$ on all the output modes after suitable rotations have been applied, here expressed by the matrix $D_{\mathrm{meas}} = \mathrm{diag} (e^{i \theta_{in}},e^{i \theta_{1}},e^{i \theta_{2}},e^{i \theta_{3}}) = (i,i,1,e^{i \theta_{3}})$. Hence the unitary transformation which transforms the three $\hat p$-squeezed modes plus the input state $(\hat a_{in},\hat a_{1},\hat a_{2},\hat a_{3})$ into the three-mode cluster state coupled with the initial state by a beam-splitter interaction reads
\be
\label{eq:ud}
U_{\mathrm{tf}} = D_{\mathrm{meas}} (U^{in,1}_{\mathrm{BS}} \otimes I^{2,3}) (I^1 \otimes U^{1,2,3}_{\mathrm{lin}}) \equiv  D_{\mathrm{meas}}  U_{\mathrm{BS}}  U^{1,2,3}_{\mathrm{lin}}
\ee
where for simplicity we have indicated with the same notation the matrix $U^{1,2,3}_{\mathrm{lin}}$, which builds the three-mode linear cluster state from three (infinitely) $p-$squeezed modes, and the same matrix times the identity on the fourth mode $I^1 \otimes U^{1,2,3}_{\mathrm{lin}}$. The matrix $U^{1,2,3}_{\mathrm{lin}}$ can be chosen as (see \ref{app:three-mode-cluster})
\ba
\label{eq:cluster-asymm}
U^{1,2,3}_{\mathrm{lin}}= \left(
\begin{array}{ccc}
 0 & -\sqrt{\frac{2}{3}} & -\frac{i}{\sqrt{3}} \\
 -\frac{i}{\sqrt{2}} & -\frac{i}{\sqrt{6}} & -\frac{1}{\sqrt{3}} \\
 -\frac{1}{\sqrt{2}} & \frac{1}{\sqrt{6}} & -\frac{i}{\sqrt{3}}
\end{array}
\right),
\ea
which, setting $\theta_3 = 0$, results in the ``desired" matrix
\be
U_{\mathrm{tf}} = 
\left(
\begin{array}{cccc}
 \frac{i}{\sqrt{2}} & 0 & \frac{1}{\sqrt{3}} & \frac{i}{\sqrt{6}} \\
 -\frac{1}{\sqrt{2}} & 0 & -\frac{i}{\sqrt{3}} & \frac{1}{\sqrt{6}} \\
 0 & -\frac{i}{\sqrt{2}} & -\frac{i}{\sqrt{6}} & -\frac{1}{\sqrt{3}} \\
 0 & -\frac{1}{\sqrt{2}} & \frac{1}{\sqrt{6}} & -\frac{i}{\sqrt{3}}
\end{array}
\right).
\ee
In order to be able to implement the matrix $U_{\mathrm{tf}}$ by means of the multi-pixel homodyne detection method, we have to find a combination of $\Delta_{\mathrm{LO}}$ and $O$ which satisfies Eq.(\ref{eq:trasf_labo_bis}) with $U_{\mathrm{th}} = U_{\mathrm{tf}}$ \footnote{Furthermore, we have to set the global phase of the local oscillator corresponding to the measurement of the $\hat p$ quadrature on all the output modes, see \ref{Appendix:matrices}.}.
We may assume here appropriate dephasing between the modes, \textit{e.g.} $\Delta^*_{\mathrm{OPO}} = \mathrm{diag}(1,1,-i,i)$. 
With the use of Eq.(\ref{eq:cod1}) we obtain 
\be
{U_{\mathrm{tf}}'}^T U_{\mathrm{tf}}' =\left(
\begin{array}{cccc}
 -\frac{i+\sqrt{2}}{\sqrt{3}} & 0 & 0 & 0 \\
 0 & \frac{i+\sqrt{2}}{\sqrt{3}} & 0 & 0 \\
 0 & 0 & \frac{-i+\sqrt{2}}{\sqrt{3}} & 0 \\
 0 & 0 & 0 & -\frac{-i+\sqrt{2}}{\sqrt{3}}
\end{array}
\right).
\ee
Among the $2^N = 16$ possible solutions for ${{\Delta_{\mathrm{LO}}}_{\mathrm{tf}}}$ and the corresponding orthogonal matrix $O_{\mathrm{tf}}$ satisfying  Eqs.(\ref{eq:cod1}, \ref{eq:cod2}) for $U_{\mathrm{th}} = U_{\mathrm{tf}}$ we can chose for instance the set
{\footnotesize
\be
\hspace{-2.3cm}  {{\Delta_{\mathrm{LO}}}_{\mathrm{tf}}}_1 = \left(
\begin{array}{cccc}
 e^{\frac{1}{2} i (\zeta +\pi )} & 0 & 0 & 0 \\
 0 & e^{\frac{i \zeta }{2}} & 0 & 0 \\
 0 & 0 & e^{-\frac{i \zeta }{2}} & 0 \\
 0 & 0 & 0 & e^{-\frac{1}{2} i (\zeta +\pi )}
\end{array}
\right) = 
\left(
\begin{array}{cccc}
 -0.3+0.95 i & 0 & 0 & 0 \\
 0 & 0.95+0.3 i & 0 & 0 \\
 0 & 0 & 0.95-0.3 i & 0 \\
 0 & 0 & 0 & -0.3-0.95 i
\end{array}
\right) \nn
\ee
}
where we have set $\zeta = \arctan (1/\sqrt{2})$. The corresponding orthogonal matrix is %${O_{\mathrm{tf}}}_1 =  \frac{1}{2}$
{\small
\be
\hspace{-2.3cm} {O_{\mathrm{tf}}}_1 =  \fr{1}{\sqrt{2}}
\hspace{-0.1cm} 
\left(
\begin{array}{cccc}
 \cos \beta & \sin \beta & -\sin \beta & -\cos \beta \\
 \sin \beta & -\cos \beta & -\cos \beta & \sin \beta \\
 -\cos \beta & -\sin \beta & -\sin \beta & -\cos \beta \\
 \sin \beta & -\cos \beta & \cos \beta & -\sin \beta
\end{array}
\right) = 
\left(
\begin{array}{cccc}
 0.67 & 0.21 & -0.21 & -0.67 \\
 0.21 & -0.67 & -0.67 & 0.21 \\
 -0.67 & -0.21 & -0.21 & -0.67 \\
 0.21 & -0.67 & 0.67 & -0.21
\end{array}
\right). \nn
\ee
}
with $\beta = \cos ^{-1}\left(\sqrt{\frac{1}{2} \left(1+\sqrt{\frac{2}{3}}\right)}\right)$.
Note that  all the modes are simultaneously measured according to the procedure described in the previous section. It is readily verified that ${O_{\mathrm{tf}}}_1 {{\Delta_{\mathrm{LO}}}_{\mathrm{tf}}}_1 G = U_{\mathrm{tf}}$. 
Hence, with the MPHD we can emulate the formation of a three-mode cluster state, the interaction of its first mode with the input mode via a beam-splitter, and the measurement of the suitable quadratures to propagate along the cluster the quantum computation. Indeed, shaping the phase of the local oscillator on the pixel modes as prescribed by the matrix ${{\Delta_{\mathrm{LO}}}_{\mathrm{tf}}}_1$ in Eq.(\ref{eq:Am1_N4again}) and choosing the digital recombination gains according to ${O_{\mathrm{tf}}}_1$ allows us to obtain a statistics for the quadrature measurement of mode $4$ as if this would contain the Fourier transform of the state in the first mode (here in a $\hat p$ squeezed state), modulo reinterpreting these result keeping into account the by-product operators. Experimentally, addressing the variety of possible input states could be realized via proper seeding of the multimode cavity with spatially and temporally shaped light. 

The same procedure also allows to implement the quadrature displacement $e^{i \hat q s}$.
This is not a symplectic operation, therefore the formalism of Ref.~\cite{Furusawa_Loock_th} cannot be used to determine the angles to measure. In reference to the scheme of Fig.\ref{fig4}, after teleporting the input state onto the second mode of the cluster state, the measurement of the quadrature $e^{-i h(\hat q)} \hat p e^{i h(\hat q)}$ on this mode allows to implement  the gate $e^{-i h(\hat q)}$, where $h(\hat q)$ is an arbitrary function of the quadrature $\hat q$~\cite{Menicucci_PRA_09}. Then, to implement the gate  $e^{i s \hat q}$ the quadrature $\hat{p}_{s \hat{q}} = e^{-i s \hat q} \hat p e^{i s \hat q} =  \hat p + s$ has to be measured. This can be achieved by measuring $\hat p$ and adding $s$ to the result. Hence, everything goes exactly as for the Fourier transform in terms of the quadratures to measure, except that we have to add ``s" to the measurement of the quadrature $\hat p$ on mode 2 of the cluster.\footnote{Note that the output state obtained by this procedure is $| \psi_{\mathrm{out}} \rangle = F e^{i \hat q s} | \psi_{\mathrm{in}} \rangle$ (apart from the by-product operators arising from the teleportation step). As can be noticed, a Fourier transform multiplies the gate $e^{i s \hat q}$~\cite{Menicucci_PRA_09, Furusawa_Loock_th}, to be interpreted  here as a by-product operator, while in the previous section it was considered the desired operation to implement.}
%, i.e. mode 3 of the cluster is projected onto \cite{Menicucci_PRA_09} 
%\be
%\label{eq:operators}
%| \psi \rangle_\mathrm{{out}} = \hat X(m_2)  F D_{h(q)} | \tilde{\psi}_{in} \rangle =  \hat X(m_2)  \tilde{f}(m_{in} ,m_1) F D_{h(q)} | \psi_{in} \rangle.
%\ee
%where $| \tilde{\psi}_{in} \rangle \equiv \hat{f}(m_{in} ,m_1) | \psi_{in} \rangle$, $\hat{f}(m_{in} ,m_1)$ are the by-product operators due to the teleportation step depending on the outcomes of the quadratures measurement $m_{in} , m_1$, $\hat X(m) = e^{- i m \hat p}$, and $\tilde{f}(m_{in} ,m_1) =  F  \hat{f}(m_{in} ,m_1) F^{\dagger} $ is the Fourier transform of the operator $\hat{f}$.
%We obtain 
%\be
%\label{eq:operators2}
%| \psi \rangle_\mathrm{{out}} =  \hat X(m_2 + s)  \tilde{f}(m_{in} ,m_1) F  | \psi_{in} \rangle.
%\ee

We have also checked that the other elementary operations of the gaussian universal set \{$e^{i \hat q s}, e^{i \hat q^2 s}, F, C_Z$\}, namely $e^{i \hat q^2 s}$, $C_Z$ may be implemented approximatively, i.e. a solution exists for matrices $U_{\mathrm{th}}$ close (in the sense of the matrix distance $d(M_1,M_2) = || M_1 - M_2 ||_F$, where $|| M ||_F = \sqrt{\sum_{i,j} |M_{i,j}|^2 }$ is the Frobenius norm) to the ones associated with these operations. The physical meaning of these approximate operations deserves however a detailed study and will be addressed elsewhere.
%-------------------------------------------------------------------------------------------------------------------------------------------------------------------------
\section{Conclusion and prospective views}

In this work we have demonstrated the possibility of measuring cluster states and implementing gaussian quantum computation in an extremely compact fashion. The method is based on the simultaneous measurement of all the (highly) squeezed optical modes in a cavity by multi-pixel homodyne detection, and on the classical post-processing of the acquired signals, which are multiplied by suitable electronic gains. This procedure requires the determination of the suitable phase shape of the local oscillator to be employed in the MPHD, as well as the gains to apply to the traces recorded in each mode. In particular, as a first example we have provided the explicit solution in terms of these experimental parameters which allows to mimic the formation of a four-mode linear cluster state. As a simple example of quantum computation, we have considered the implementation of a Fourier transform on an input mode.  

It is important to stress that by the Gottesmann-Knill theorem~\cite{Menicucci_PRA_09, Mari_eisert} it is known that all the results of manipulations with gaussian elements, such as the squeezed states and homodyne detection discussed in this work, can be efficiently simulated with a classical computer. In order to perform operations which outperform classical processing capability at least a non-gaussian operation is necessary. The inclusion in the experimental apparatus of a photon counter which detects the number of photons in one mode ~\cite{Christine} could lead to the extension of the class of the accessible operations, yielding to the implementation of non-gaussian unitaries.

Finally, we remark that in our set-up the state preparation and the computation are performed in a single step, yielding hence a \emph{quantum depth} equal to $1$ - the quantum depth being indeed defined as the number of measurement steps that a computation task requires, when trying to do as many measurements as possible at the same time. This may provide an advantage with respect to the circuit model for performing the same operation~\cite{Elham}.%\footnote{In the circuit model, the depth is defined as the minimal number of time steps needed to implement a certain task when doing as many gates as possible at the same time in each step.}. 
%'depth complexity' is just about how many time steps it takes to do a computation. For MBQC this is how many measurement steps, when trying to do as many as possible at the same time. For the circuit based model, it is.

%-------------------------------------------------------------------------------------------------------------------------------------------------------------------------
\section{Acknowledgments}

We acknowledge useful discussions with J.-F. Morizur and D. Markham. The research is supported by the ERC starting grant FREQUAM, and by the ANR project NAMOCS. C. Fabre is a member of the Institut Universitaire de France. 

%%%%%%%%%%%%%%%%%%%%%%%%%%%%%%%%%%%%%%%%%%%%%%%%%%%%%%%%%%%%%%%%%%%%%%%%%%%%%%%%%%%

\appendix
%{Supplementary informations}
%-------------------------------------------------------------------------------------------------------------------------------------------------------------------------

\se{Modelization of the operations performed on the modes}
\label{Appendix:matrices}

%-------------------------------------------------------------------------------------------------------------------------------------------------------------------------
\sse{Basis and notations}

We introduce here the notations and conventions that we are going to use. The local bosonic operator $\hat{a}(\rho)$ satisfies
\be
[\hat{a}(\rho),\hat{a}^{\dagger}(\rho') ] = \delta(\rho - \rho')
\ee
and is related to the local quadrature operators by
\ba
\label{eq:basis1}
\hat{q} (\rho) &=& \hat{a}^{\dagger}(\rho) + \hat{a}(\rho) \nn \\
\hat{p} (\rho) &=& i \lt \hat{a}^{\dagger}(\rho) - \hat{a}(\rho) \rt.
\ea
%and
%\ba
%\label{eq:basis2}
%\hat{a}_{k} &=& \fr{1}{2} (\hat{x}_{k} + i \hat{p}_{k}) \nn \\
%\hat{a}_{k}^\dagger &=& \fr{1}{2} (\hat{x}_{k} - i \hat{p}_{k}) 
%\ea
so that $[\hat{q}(\rho),\hat{p}(\rho') ] =  2 i \delta(\rho - \rho')$.
Given an orthonormal basis of wave functions $u_k(\rho)$ satisfying the orthogonality relation
\ba
\label{eq:orthogonality}
 \int d  \rho \, u^*_k  (\rho) u_l  (\rho) = \delta_{k,l} 
% \int \hspace{-0.2cm} \int_{\rho \in \R^2} d^2  \rho \, u^*_k  (\rho,z) u_l  (\rho,z) = \delta_{k,l} 
%\sum_k u_k  (\rho,z) u^*_k  (\vec{\rho'},z)  = \delta(\rho - \rho'),
\ea
the local bosonic operator can be decomposed as
\be
\label{eq:def_modo2}
\hat{a}(\rho) = \sum_{k=0}^{\infty}  \hat{a}_{u_k} u_k  (\rho) 
\ee
with
\be
\label{eq:def_modo}
%\hat{a}_{u_k} =  \int \hspace{-0.2cm}\int_{\rho \in \R^2} d^2 \rho \hat{a}(\rho,z) u^*_k  (\rho,z).
\hat{a}_{u_k} =  \int d \rho \hat{a}(\rho) u^*_k  (\rho).
\ee
%
%Note that in principle the summation is to be taken with $1 \leq k \leq \infty$, unless for special cases in which a finite basis may be taken.

%-------------------------------------------------------------------------------------------------------------------------------------------------------------------------
\sse{Multimode homodyne detection}
\label{sse:mhd}

We procede now with the modelization of the change of basis which the optical modes undergo as an effect of the multi-pixel homodyne detection, as reported in Ref.\cite{Jeff_thesis}.
The beams splitter mixes the input modes with the local oscillator, yielding the modes
\ba
\hat{a}_{A,u_k} = \fr{1}{\sqrt{2}} \lt  \hat{a}_{\mathrm{LO},f_k}  + \hat{a}_{u_k}  \rt \nn \\
\hat{a}_{B,u_k} = \fr{1}{\sqrt{2}} \lt  \hat{a}_{\mathrm{LO},f_k}  - \hat{a}_{u_k}  \rt
\ea
with
\ba
\label{eq:scomposizione_A_field}
&& \hat{a}_A(\rho) =  \sum_k   \hat{a}_{A,u_k} u_k (\rho) \nn \\
&&  \hat{a}_{A,u_k} =   \int_{\rho \in \R} \hspace{-0.4cm} d  \rho \, u^*_k  (\rho) \hat{a}_A(\rho).
\ea
Then, light is propagated to the pixels. Let us introduce the output intensities on each pixel
\ba
\hat{i}_{A, i} =  \int_{\rho \in S_{i}} d  \rho \, \hat{a}^\dagger_A(\rho) \hat{a}_A(\rho) \nn \\
\hat{i}_{B, i} = \int_{\rho \in S_{i}} d  \rho \, \hat{a}^\dagger_B(\rho) \hat{a}_B(\rho).
\ea
Taking the differences of the intensity signals between couples of corresponding pixels gives
\ba
\hat{s}_{i } &=& \hat{i}_{A,i} - \hat{i}_{B,i} \nn \\
&=& \fr{1}{2} \sum_{k,l} \lqu (\hat{a}^\dagger_{\mathrm{LO},u_k} + \hat{a}^\dagger_{u_k}) (\hat{a}_{\mathrm{LO},u_l} + \hat{a}_{u_l}) - (\hat{a}^\dagger_{\mathrm{LO},u_k} - \hat{a}^\dagger_{u_k}) (\hat{a}_{\mathrm{LO},u_l} - \hat{a}_{u_l}) \rqu \nn \\
& &  \times  \int_{\rho\in S_{i}} \hspace{-0.4cm} d  \rho \, u^*_k  (\rho) u_l  (\rho) \nn \\
&=& \sum_{k,l} \lt \hat{a}^\dagger_{\mathrm{LO},u_k}\hat{a}_{u_l} + \hat{a}^\dagger_{u_k}\hat{a}_{\mathrm{LO},u_l} \rt \int_{\rho\in S_{i}} \hspace{-0.4cm} d  \rho \, u^*_k  (\rho) u_l  (\rho)
\ea
Now we assume that the local oscillator is in an intense coherent state in a certain mode that we will indicate by $\hat{a}_{\mathrm{LO},u_{\mathrm{LO}}}$ ($u_{\mathrm{LO}}$ is indeed one of the modes of the basis $\lgr u_i \rgr$), and we keep only the terms implying such a mode. Furthermore, we use %$\delta \hat{a}^\dagger_{\mathrm{LO},u_{\mathrm{LO}}}$, using 
$\hat{a}^\dagger_{\mathrm{LO},u_{\mathrm{LO}}} \simeq \langle \hat{a}^\dagger_{\mathrm{LO},u_{\mathrm{LO}}} \rangle = \alpha_0^*$, being $\alpha_0 \gg \langle \hat{a}^\dagger_{u_k} \rangle \,  \forall \, k$. With this we obtain 
\ba
\label{eq:detected_signal}
\hat{s}_{i } &\simeq& \sum_{l}  \hat{a}^\dagger_{\mathrm{LO},u_{\mathrm{LO}}}\hat{a}_{u_l} \int_{\rho\in S_{i}} \hspace{-0.4cm} d  \rho \, u_{\mathrm{LO}}^* (\rho) u_l  (\rho) + 
                                    \sum_{k} \hat{a}^\dagger_{u_k}\hat{a}_{\mathrm{LO},u_{\mathrm{LO}}}  \int_{\rho\in S_{i}} \hspace{-0.4cm} d  \rho \, u^*_k (\rho) u_{\mathrm{LO}}  (\rho) \nn \\
                     &\simeq& \sum_{k} \lt  \alpha^*_0 \hat{a}_{u_k} \int_{\rho\in S_{i}} \hspace{-0.4cm} d  \rho \, u_{\mathrm{LO}}^* (\rho) u_k  (\rho) +
                                              \alpha_0 \hat{a}^\dagger_{u_k} \int_{\rho\in S_{i}} \hspace{-0.4cm} d \rho \, u^*_k (\rho) u_{\mathrm{LO}}  (\rho) \rt .                                                  
\ea
%We now define the pixel modes as
According to the definition in Eq.(\ref{eq:def_modo}) and to Eq.(\ref{eq:pixel_modes}) we can write
\ba
\label{eq:sb}
\hat{a}_{v_{i}} =   \int_{\rho \in \R} d \rho \hat{a}(\rho) v_{i}^*  (\rho) = \kappa_i  \int_{\rho\in S_{i}} \hspace{-0.4cm} d \rho   \hat{a}(\rho)  u_{\mathrm{LO}}^*   (\rho).
\ea
Substituting Eq.(\ref{eq:def_modo2}) in Eq.(\ref{eq:sb}) we obtain
\ba
\label{eq:ac}
\hat{a}_{v_{i}} =  \kappa_i \sum_j  \int_{\rho\in S_{i}} \hspace{-0.4cm} d \rho  u_{\mathrm{LO}}^*(\rho)  u_j(\rho) \hat{a}_{u_j}  = \sum_j {U_T}_{i,j} \hat{a}_{u_j},
\ea
where we have defined the matrix realizing the change of bases
\be
\label{eq:coeffs_c_ij}
U_{T \, i, j} = \kappa_i  \int_{\rho\in S_i} \hspace{-0.4cm} d \rho  u_{\mathrm{LO}}^*(\rho)  u_j(\rho).
\ee
Remark that the matrix $U_T$ is not necessarily square nor has finite dimension.  
In terms of the pixel modes defined in Eq.(\ref{eq:pixel_modes}) Equation (\ref{eq:detected_signal}) becomes
\ba
\hat{s}_{i } &=& \lt  \alpha^{'*}_0 \hat{a}_{v_{i}} + \alpha'_0 \hat{a}^\dagger_{v_{i}}  \rt \nn \\
&=&   |\alpha_0'| \fr{1}{2} \lt e^{i \varphi_0} (\hat{q}_{v_{i}} + i \hat{p}_{v_{i}})   +   e^{-i \varphi_0} (\hat{q}_{v_{i}} - i \hat{p}_{v_{i}})  \rt  \nn \\
&=& |\alpha_0'| (\hat{q}_{v_{i}} \cos \varphi_0 - \hat{p}_{v_{i}} \sin \varphi_0 )  = |\alpha_0'| \hat{q}^{\varphi_0}_{v_{i}},
\ea 
where we have set $\alpha_0' = |\alpha_0| e^{-i \varphi_0}/\kappa$. We see that the measurement of the quadrature $\hat p$ on each output mode requires setting the phase of the local oscillator to $\varphi_0 = 3 \pi /2$.

%------------------------------------------------------------------------------------------------------------------
\sse{Signal recombination} 
\label{Signal recombination}

The pixel modes can be digitally multiplied by real gains and recombined, leading to the modes $v^{''}_{i} (\rho) = \sum_{j} g^i_{j} v'_{j} (\rho)$. Since the output functions $v^{''}_{i} (\rho)$ should form an orthonormal basis, we are restricted to gains which are elements of an orthogonal matrix, i.e.  $O_{i,j} = g^i_{j}$. Then the corresponding annihilation operators are
\ba
\hat{a}_{v^{''}_{i} } = \sum_{j} O_{i,j} \hat{a}_{v'_{j}},
\ea
and we can obtain information on a quadrature as
\ba
\label{eq:step0}
\hat{s}^{''}_{i} &=& \sum_{j} O_{i,j} \hat{s}'_{j } = \lt  \alpha^{'*}_0 \sum_{j} O_{i,j}  \hat{a}_{v'_{j}} + \alpha_0' \sum_{j} O_{i,j} \hat{a}^\dagger_{v'_{j}} \rt  \nn \\
&=& \lt  \alpha^{'*}_0 \hat{a}_{v^{''}_i} + \alpha_0' \hat{a}^\dagger_{v^{''}_i}  \rt  \nn \\
&=& |\alpha_0'| (\hat{q}_{v^{''}_{i}} \cos \varphi_0 - \hat{p}_{v^{''}_{i}} \sin  \varphi_0  ) \nn \\
&=& |\alpha_0'| (\hat{q}_{v^{'}_{i}} \cos (\varphi_0 +\varphi_i) - \hat{p}_{v^{'}_{i}} \sin  (\varphi_0 + \varphi_i) )
\ea 

%-------------------------------------------------------------------------------------------------------------------------------------------------------------------------
\section{Three-mode linear cluster state matrix}
\label{app:three-mode-cluster}

In this appendix we derive the matrix $U^{1,2,3}_{\mathrm{lin}}$ which builds-up the $3$-component linear cluster state. For a general cluster state identified by an adjacency matrix $V$ the corresponding unitary matrix which brings from the squeezed modes to the given cluster state can be chosen as \emph{any} possible matrix satisfying~\cite{Loock_unitary}
\ba
\label{eq:sol_cluster}
& U = X + i Y; \nn \\
& Y - V X = 0 \nn \\
& X X^T + Y Y^T = \mathcal{I}; \nn \\
& X^TY = Y^T X; \, \, \, X Y^T = Y X^T.
\ea
Setting $A = X X^T$ the previous equations leads to the linear system 
\be
\label{eq:linear_sys}
V A V = \mathcal{I} - A.
\ee
For a three-mode linear cluster state, the adjacency matrix is given by
\be
\label{eq:V}
V  = 
\left( 
\begin{array}{cccccccc}
0 & 1  & 0  \\
1  & 0 & 1 \\
0 &  1 & 0 \\
\end{array}
\right).
\ee
The linear system in Eq.(\ref{eq:linear_sys}) yields in this case the solution 
\be
\label{eq:A}
A = \left(
\begin{array}{ccc}
 \frac{2}{3} & 0 & -\frac{1}{3} \\
 0 & \frac{1}{3} & 0 \\
 -\frac{1}{3} & 0 & \frac{2}{3}
\end{array}
\right).
\ee
From Eq.(\ref{eq:A}) we easily obtain the symmetric solution $X_s$ such that $X_s^2 = A$, and the corresponding $U_s$ and $V_s$ satisfying Eq.(\ref{eq:sol_cluster}) 
\ba
\label{eq:X}
\hspace{-2.1cm}
X_s = \left(
\begin{array}{ccc}
 \frac{\left(3+\sqrt{3}\right)}{6}  & 0 & \frac{\left(-3+\sqrt{3}\right)}{6} \\
 0 & \frac{1}{\sqrt{3}} & 0 \\
 \frac{\left(-3+\sqrt{3}\right)}{6}  & 0 & \frac{\left(3+\sqrt{3}\right)}{6} 
\end{array}
\right);  \,
Y_s = \left(
\begin{array}{ccc}
 0 & \frac{1}{\sqrt{3}} & 0 \\
 \frac{1}{\sqrt{3}} & 0 & \frac{1}{\sqrt{3}} \\
 0 & \frac{1}{\sqrt{3}} & 0
\end{array}
\right);  \, 
U_{s} = \left(
\begin{array}{ccc}
 \frac{\left(3+\sqrt{3}\right)}{6}  & \frac{i}{\sqrt{3}} & \frac{\left(-3+\sqrt{3}\right)}{6}  \\
 \frac{i}{\sqrt{3}} & \frac{1}{\sqrt{3}} & \frac{i}{\sqrt{3}} \\
 \frac{\left(-3+\sqrt{3}\right)}{6}  & \frac{i}{\sqrt{3}} & \frac{\left(3+\sqrt{3}\right)}{6} 
\end{array} \nn
\right).
\ea
From this symmetric solution, all the other solutions can be obtained. Indeed, it is easy to see from Eq.(\ref{eq:sol_cluster}) that if $X_s$ is a solution, also $X = X_s \mathcal{O}$ is a solution for any real orthogonal matrix $\mathcal{O}$. Hence, any matrix $U^{1,2,3}_{\mathrm{lin}} = X + i Y = X_s \mathcal{O} + i V X_s \mathcal{O} = (\mathcal{I} + i V) X_s \mathcal{O} = U_s \mathcal{O}$ can be used to build a cluster state. 
In order to explicit the free parameters induced by this rotation, we will use the Euler parameterization
\be
\mathcal{O} =  \left(
\begin{array}{ccc}
 \cos {\psi} & \sin {\psi} & 0 \\
- \sin {\psi} & \cos {\psi} &0 \\
 0 & 0 & 1 \\
\end{array}
\right)
\left(
\begin{array}{ccc}
 \cos {\theta} & 0 & \sin {\theta} \\
0 & 1 &0 \\
 - \sin {\theta} & 0 & \cos {\theta}  \\
\end{array}
\right)
\left(
\begin{array}{ccc}
 \cos {\phi} & \sin {\phi} & 0 \\
- \sin {\phi} & \cos {\phi} &0 \\
 0 & 0 & 1 \\
\end{array}
\right).
\ee
With this we obtain $U^{1,2,3}_{\mathrm{lin}} = U_s \mathcal{O}$. An example is provided in Eq.(\ref{eq:cluster-asymm}) of the main text.

%\end{widetext}
%-------------------------------------------------------------------------------------------------------------------------------------------------------------------------

\end{document}